\documentclass[aps,prb,showpacs,twocolumn,preprintnumbers,amsmath,amssymb,superscriptaddress,floatfix]{revtex4}
\usepackage[dvips]{graphicx}
\usepackage{mathptmx}
\usepackage{verbatim}

\begin{document}
\title{Non-local exchange effects in zigzag edge magnetism of neutral graphene nanoribbons}
\author{Jeil Jung} \email{jeil@physics.utexas.edu}
\affiliation{Department of Physics, University of Texas at Austin, Austin, USA}

\begin{abstract}
We study the role of non-locality of exchange in a neutral zigzag graphene nanoribbon
within the $\pi$-orbital unrestricted Hartree-Fock approximation. 
Within this theory we find that the magnetic features are further stabilized for both the intra-edge 
and inter-edge exchange compared to mean field theories of zigzag ribbon based on local exchange 
(like the Hubbard model or the ab-initio Local Density Approximation).  
The inter-edge exchange produces an enhancement of the band-gap of the magnetic ground-state solutions.  
The effect of this enhanced exchange on the edge states can not be satisfactorily 
achieved by a local interaction with renormalized parameters.
\end{abstract}

\maketitle

\section{Introduction}
Zigzag terminated graphene ribbons have recently attracted a lot of attention both by theorists
\cite{fujita, nakada, waka, ezawa, brey,  hikihara,leehosik,kusakabe,son_gap,son_half,pisani,yazyev, joaquin, pohang,  superexchange, rhim, white, sandler, sudipta, doping,magnetoelectric,sudipta1,xu,kuntsmann}  
and experimentalists \cite{niimi,kobayashi,han,chem_ribbon,etching,jarillo,rice,dresselhaus,zettl,chuvilin,ye,enoki,joly,crommie}    
partly due to the surge in popularity of graphene after seminal transport experiments. \cite{novoselov,pkim,graphenereviews}
The unique properties of zigzag edge localized states 
has prompted using zigzag terminated nanoribbons as a testbed for 
proposals of magnetism in graphitic systems \cite{coey}
or for illustrating exotic physics related to the bulk 
topology of single sheet of multilayer graphene in presence of strong spin-orbit coupling or magnetic fields. \cite{kane,arikawa,qiao,zhang}
Recent experiments have found traces of localized spin 
polarization at zigzag edges of graphene \cite{enoki,joly,crommie}
in agreement with theories predicting spin polarization at zigzag edges in presence of electron
interactions. \cite{fujita,kusakabe,son_gap,pisani} 
The agreement of the qualitative features between solutions predicted by
different methods such as the Hubbard model and DFT with different
semi-local approximations for the energy functionals point towards
a robust underlying property of these systems for which the magnetic solutions
are not very sensitive to the specific details of electron interaction.
Considering that magnetic ordering is a long-ranged phenomena one natural question that follows
is how much physics we are missing by an inadequate treatment of non-local exchange in our solutions.
The Hubbard model used in several previous papers \cite{joaquin,yazyev1,joaquin1,julio}
relies on a strictly local onsite interaction term while the
common approximations of the energy functionals in DFT \cite{perdew,pbe}
are modeled from local or semi-local exchange and correlation energies \cite{ceperley} of the homogeneous
electron gas. 
In both cases the description of the exchange hole tends to be excessively short ranged  
in addition to uncertainties associated with self-interaction errors in the case of DFT.
In this work we revisit the problem of edge magnetization, predicted by mean field
theories in zigzag nanoribbons, within Unrestricted Hartree Fock (UHF) approximation
with non-local exchange, that does not suffer from the shortcomings of
the previous approximations, and study systematically different truncation ranges 
for the electron interaction.

We find that a non-local interaction enhances the gaps in the band structure,
a feature that can be traced back to both the inter-edge tunneling term and intra-edge magnetic order.
As we discuss below models with short-range interaction are unable to mimic 
the large size of the gap in a consistent way although the 
solutions remain similar in many aspects.

We start in section II introducing the Hartree-Fock theory in zigzag ribbons and cast it in the form 
of a two dimensional edge state bands model  \cite{joaquin,superexchange} 
extending beyond the formulation based on the Hubbard hamiltonian.  
In section III we show within this framework how the non-local exchange 
modifies the effective hamiltonian matrix elements when we extend the onsite interaction Hubbard model
to include farther neighbor interaction terms.
We then move on to discuss in section IV the localization properties of the edge state wave functions along 
the direction of the ribbon discussing the dependence of the intra-edge band-gap as a function of the electron 
interaction strength. 
Subsequently in section V we assess the impact of non-local exchange on the energetics of the inter-edge 
antiferromagnetic ground state. We finally close the paper with a summary and conclusions section.

\section{Two bands Hartree-Fock theory in zigzag ribbons}
A simple and yet fairly accurate way to study edge magnetism in a zig-zag ribbon
in a mean field theory is to calculate, for each $k$-point, the effect of interaction on 
the edge state wave functions not allowing edge and bulk to mix their wave 
functions\cite{joaquin,superexchange}.
These edge state wave functions (see appendix for more details) become exponentially 
localized at the edges whenever $2\pi/3a + 1 / W \leq  \left|k \right| $ for sufficiently wide ribbons
\cite{brey, malysheva} where the ribbon width is $W = \sqrt{3} a N / 2$
and $N$ is the number of atom pairs in the unit cell across the ribbon and the lattice
constant is $a = 2.46 \AA$.
We label in an abbreviated  notation the ribbons edge states as
$\left| k - \right>$ and $\left| k + \right>$, antisymmetric and symmetric
wave functions across the ribbon, that can be symmetrically and antisymmetrically
combined to obtain basis functions that are
mostly centered either at the (L)eft or (R)ight edge in the ribbon. \cite{superexchange}
We denote the respective $k$-dependent wave function amplitude at each lattice site
$l$ in the unit cell as $L_{kl}$ and $R_{kl}$.
We can use this new basis to represent the Hamiltonians as a two by two matrix
for each spin $\sigma = \uparrow / \downarrow$
\begin{equation}
\label{tbhamil}
H_{\sigma} \left( k \right) =
\left( \begin{array}{cc}
  H_{LL, \, \sigma}\left( k \right)  &  H_{LR, \, \sigma}\left( k \right) \\
  H_{RL, \, \sigma} \left( k \right) & H_{RR, \, \sigma} \left( k \right)
\end{array} \right).
\end{equation}
The tight-binding band term Hamiltonian is described by a simple nearest neighbor
hopping term $\gamma_0 = -2.6 eV$. 
In this LR basis representation the Hamiltonian would consist 
of a tunneling term mixing states located at both edges in the ribbon for each spin
\begin{equation}
\label{hamiltb}
H^{TB} \left( k \right) =
        \left(
        \begin{array}{cc}
  0    &   t_{TB} \left( k \right)      \\
  t_{TB} \left( k \right)    &   0
        \end{array}
        \right)
\end{equation}
where $t_{TB}\left( k \right)$ is the energy dispersion for the tight-binding
edge conduction band ( i.e., the basis we use consists of two eigenstates of $H^{TB}\left( k \right) )$.
For a neutral ribbon the electrostatic Hartree term cancels with the positive
background charge and contributes only with a constant term $C_0$ in the diagonal
elements that can be chosen at our convenience.
Then the interaction term of the Hartree-Fock hamitonian projected on the two-bands LR 
basis effectively reduces to an exchange contribution on both the diagonal and off diagonal 
matrix elements
\begin{eqnarray}
\label{lhamil}
H_{LL, \, \sigma} \left(k \right) &=& C_0
- \sum_{l l^{\prime}}
L_{k l} L_{k l^{\prime}}
F_{X, \, \sigma}^{l l^{\prime}}\left( k \right)
\\
\label{lrhamil}
H_{LR, \, \sigma} \left(k \right)
&=&  t_{TB} \left( k \right)
- \sum_{l l^{\prime}} L_{k l} R_{k l^{\prime}}
F_{X, \, \sigma}^{l l^{\prime}}\left( k \right)
\\
\label{rhamil}
H_{RR, \, \sigma} \left(k \right) &=& C_0
- \sum_{l l^{\prime}}
R_{k l} R_{k l^{\prime}}
F_{X, \sigma}^{l l^{\prime}}\left( k \right)
\end{eqnarray}
where we have defined
\begin{eqnarray}
F_{X, \, \sigma}^{l l^{\prime}} \left( k \right) &=&
\sum_{k^{\prime}}
U^{l l^{\prime}}_{X}
\left( k^{\prime} - k \right)
\left< n_{k^{\prime} l l^{\prime}, \, \sigma} \right>
\end{eqnarray}
where $n_{k^{\prime} l l^{\prime}, \, \sigma}$ is the density matrix
in the lattice Bloch states and $U_{X}^{l l^{\prime}} \left( k^{\prime} - k \right)$
represents exchange Coulomb integrals whose explicit expressions are 
presented in the appendix.
The Coulomb integrals between the $\pi$-orbitals located at different sites
can be approximated in Hartree atomic units by effective terms of the form\cite{zarea, gogolin} 
\begin{eqnarray}
V_{\rm eff}\left( d \right) = \frac{1}{ \epsilon_r \sqrt{ a_o^2 + d^2} }
\label{coulint}
\end{eqnarray}
where $d$ is the distance between lattice sites and the bonding radius of the carbon 
atoms $a_o = a/\left(2 \sqrt{3}\right)$ account for a small damping of the interaction
due to the shape and finite spreading of the wave function around the lattice.
The onsite repulsion term remains unknown and is commonly bracketed between
$U = 2 eV$ and $U = 6 eV$.  \cite{alicea,yazyev1,bhowmick,wunsch,monolayer_hf,joaquin1,julio}
In our calculation we used a relatively small value of $U = 2.5 eV$,
considering that the onsite repulsion strength of $U = 2 eV$ in the Hubbard model where the Coulomb tail
is neglected is enough to reproduce the LDA band structures \cite{joaquin}.
We include in our calculations an effective screening of $\epsilon_r = 4$ resulting in
a global damping of the interaction strength by a factor 1/4 to account for possible
effects of dielectric screening (usually $\epsilon_r = 2.5$ in $Si O_2$) and
additional screening effects due to $sp_2$ orbitals that are neglected in our approximation.
The range of the Coulomb interaction is truncated to up 16 nearest neighbors.
With the above prescriptions we obtain for the ground state of the
neutral system an intra-edge interaction gap near $ka \sim \pi$ of $1.4 eV$
midway between the B3LYP prediction\cite{pisani} of $2.2 eV$
and the LDA prediction\cite{son_gap} of $0.5 eV$.
We write the two-bands effective Hamiltonian in the following way: 
\begin{eqnarray}
\label{h3term}
H_{\sigma} \left( k \right) &=& \Delta^{AF} \left( k \right) \sigma \tau_{z}+ \Delta^{F} \left( k \right) \sigma I 
+ t_{\sigma} \left( k \right) \tau_{x},
\end{eqnarray}
where the pauli matrices $\tau_\mu$ and $I$ are acting in the space of left and right edges and $\sigma$ 
is $+(-)$ for up(down) spin.
The renormalized inter-edge tunneling $t_{\sigma} \left( k \right)$ is given by the bare hopping plus a spin dependent enhancement from the exchange interaction.  
The intra-edge exchange potentials $\Delta^{AF} \left( k \right)$ and $\Delta^{F} \left( k \right)$ 
correspond to a particular choice of $\Delta^{AF}_{\uparrow / \downarrow} \left( k \right)$
or $\Delta^{F}_{\uparrow/ \downarrow} \left( k \right) $
such that they are positive quantities at $k \sim \pi/a$. 
Written in terms of the matrix elements defined in Eq.\ (\ref{lhamil},\ref{lrhamil},\ref{rhamil}) for each momentum $k$ they are:
\begin{eqnarray}
\label{tdef}
t_{\sigma} \left( k \right)   &=& H_{LR, \, \sigma} \left( k \right) \\
\label{deltadef}
\Delta^{AF}_{\sigma} \left( k \right) &=& \frac{1}{2}
\left( H_{LL, \, \sigma} \left( k \right) - H_{RR, \, \sigma} \left( k \right) \right) \\
\label{ddef}
\Delta^{F}_{\sigma} \left( k \right)  &=& \frac{1}{2}
\left( H_{LL, \, \sigma}
+ H_{RR, \, \sigma} \right) \left( k \right).
\end{eqnarray}
Note that we have assumed a collinear spin arrangement and there is no spin-mixing term.  
Therefore the $4\times4$ Hamiltonian matrix
is reduced to two block diagonal sub-matrices in Eq.\ (\ref{h3term}) for each spin sub-space. 
The energy dispersions associated with the edge states of the
two-band Hamiltonian can be written as
\begin{eqnarray}
E^{\pm}_{\sigma} \left( k \right)
= \sigma\Delta^{F}\left( k \right) \pm \sqrt{{\Delta^{AF}}^{2} \left( k \right) + t^2_{\sigma} \left( k \right)}.
\end{eqnarray}
In the AF case
the term $\Delta^{F} \left( k \right)$ in equation (\ref{h3term})
amounts to an edge-independent value for both up and down spin and can be set to zero.
We notice that the exchange spin splitting terms given by
the diagonal elements $\Delta^{AF} \left( k \right)$ have opposite signs for left and right edge states, 
and a global sign reversal when we consider the opposite spin Hamiltonian.
On the other hand, in ferromagnetic self-consistent solutions we get a nonzero net spin polarization 
in the sample due to edge electrons with spins pointing in the same direction.
In this case because $\Delta^{AF}_{\sigma} = 0$
the intra-edge exchange energy gain included in
$\Delta^{F}_{\sigma} \left( k \right)$
is reflected in the overall shift of the band energy for each $k$.
The opposite spin state is shifted in an equal and
opposite direction and the bands split in an amount of $2\Delta^{F} \left( k \right)$
at each $k$-point.

\section{Non-local exchange and the effective Hamiltonian matrix elements}
The non-locality in exchange appear when the Coulomb interaction range is extended 
beyond the onsite repulsion and is most easily explored comparing the Hartree-Fock 
Hamiltonian matrix elements of equations (\ref{lhamil},\ref{lrhamil},\ref{rhamil})
with those we obtain using the Hubbard onsite repulsion model.
We have used the value of $U = 2 eV$ for our reference Hubbard model calculations
following the convention of previous works \cite{joaquin, superexchange} to 
match the LDA band gaps.
In Fig. 1 we show a comparison of the band structures in the HF and Hubbard model
for the lower energy AF and higher energy F configurations.
\begin{figure}[htbp]
\label{ex_distr}
\begin{center}
\begin{tabular}{cc}
\resizebox{43mm}{!}{\includegraphics{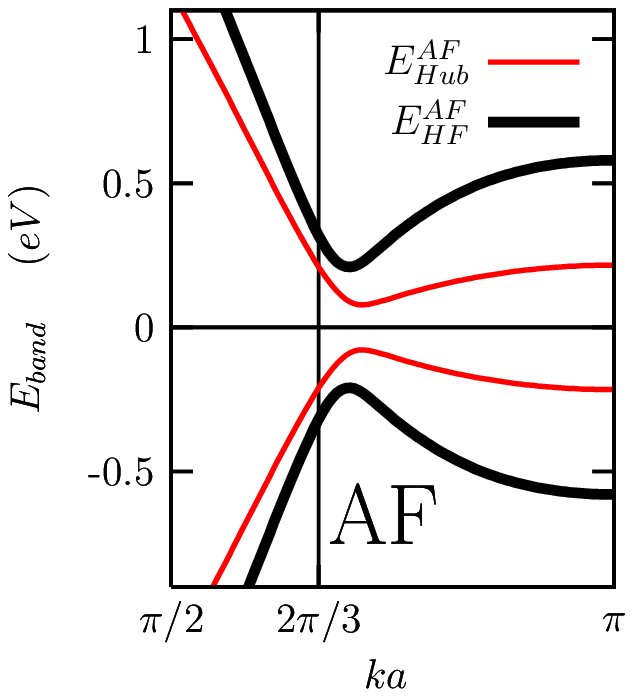}}  &  \resizebox{37.4mm}{!}{\includegraphics{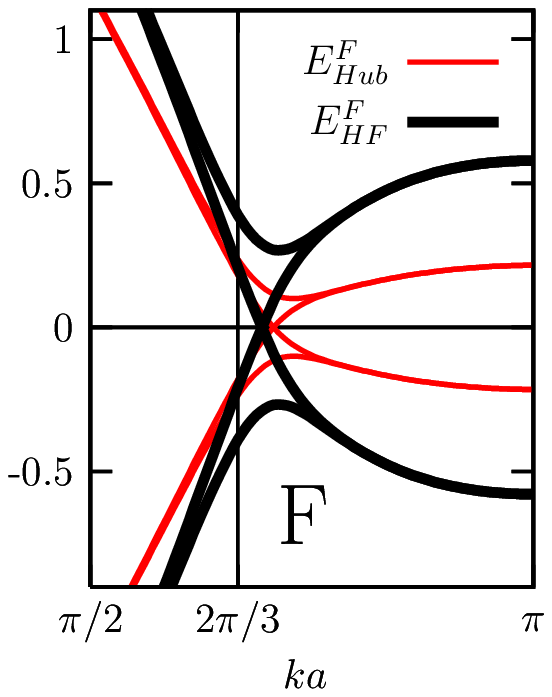}} \\
\end{tabular}
\caption{
(Color online)
Comparison of HF and Hubbard model band structures for the lower 
energy AF (left panel) and meta-stable higher energy F (right panel) spin configurations 
for a nanoribbon with $N=20$ carbon atom pairs in the unit cell.
In the HF solution we find enhanced band-gap openings due to the increase in strength
of both intra-edge exchange manifested near $ka \sim \pi$ and inter-edge tunneling 
that become more significant as we move away from this point.
}
\label{figex}
\end{center}
\end{figure}
The effects of non-local exchange in the band structure can be understood more clearly
separating their contributions for the intra-edge exchange potentials corresponding to the
diagonal elements of the effective Hamiltonian obtained through equations
(\ref{lhamil},\ref{rhamil},\ref{deltadef}) and the off diagonal inter-edge tunneling given in equations
(\ref{lrhamil},\ref{tdef}).
We discuss our comparison only for AF matrix elements because the effective matrix 
elements for F solutions remain qualitatively identical.
These matrix elements are represented in Fig. (2) for different choices in the
truncation of the interaction ranges for the Coulomb integrals in equation (\ref{coulint}).
Since the left and right basis functions are strictly localized on one of the sublattices
the contributions to the diagonal term $\Delta^{AF} \left( k \right)$ are due to
interaction couplings within the same sublattice, therefore these terms contain information
of exchange within the same edge atoms.
\begin{figure}[htb]
\label{longranged}
\begin{center}
\begin{tabular}{cc}
\resizebox{41.5mm}{!}{\includegraphics{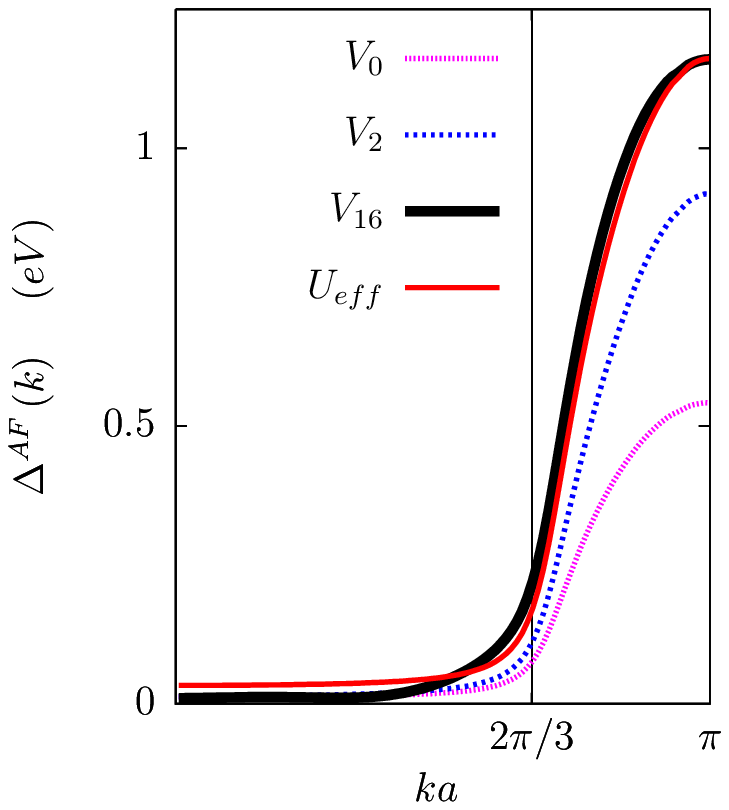}}  $\,\,\,$ & $\,\,\,$
\resizebox{39mm}{!}{\includegraphics{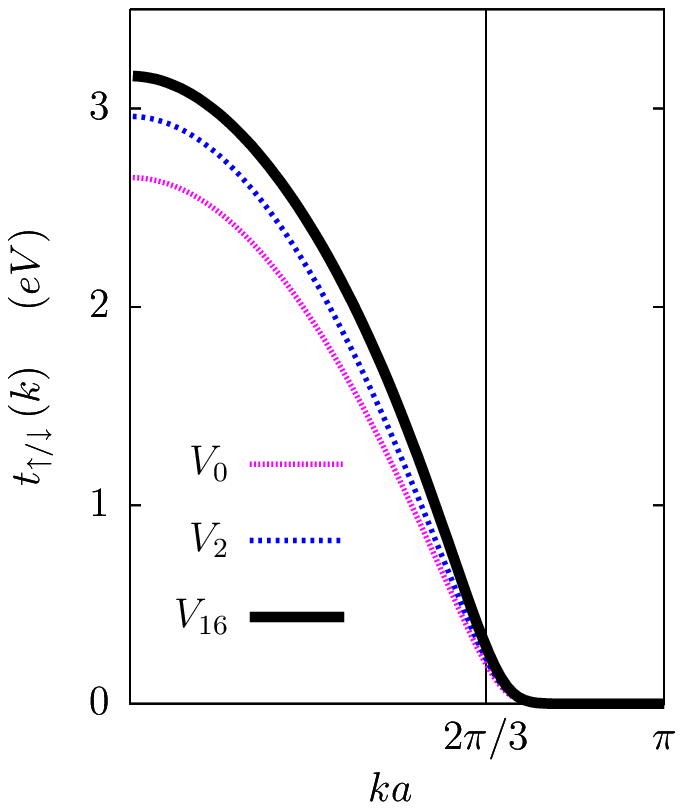}}  \\
    \end{tabular}
\caption{(Color online)
Enhancement of the matrix elements of the effective hamiltonian defined in equations 
(\ref{tdef},\ref{deltadef},\ref{ddef}) evaluated in the HF approximation on a ribbon with 
$N=20$. The subscripts in the label indicate the number of nearest neighbors considered 
in the truncated range of the effective Coulomb interaction.
We do not represent the results for F configuration because the behavior is similar
to what is found for AF solutions.
{\em Left Panel:} 
Values of the intra-edge exchange potential $\Delta^{AF} \left( k \right)$ obtained in the 
HF approximation for different interaction ranges.
With an appropriate choice of an effective $U_{\rm eff} = 5.3 eV$ in the Hubbard model 
we can match $\Delta^{AF} \left( \pi/a \right)$ of the HF calculation 
for most of the region in the Brillouin zone where the wave functions are edge localized.  
The red curve represents the effective Hubbard model that agrees well with the
HF calculation in almost the entire range of edge states zone $\left| k \right|  \leq 2 \pi /3a + 1/W$.
{\em Right Panel:}
Enhancement of tunneling amplitudes $t_{\uparrow / \downarrow} \left( k \right)$ for
AF solutions in presence of long-ranged interaction representing the greater coherence
between wave functions located at each sublattice, and therefore each edge in the ribbon. 
For onsite interactions the tunneling has the same form as the tight-binding conduction 
band dispersion.
}
\end{center}
\end{figure}
One interesting feature we observe is that the function $\Delta^{AF} \left( k \right)$ obtained for
the fully non-local calculation can be effectively reproduced in most of the edge states zone
of the Brillouin zone ($\left| k \right| \leq 2 \pi /3a + 1/W$) by the Hubbard Hamiltonian with a 
larger effective onsite repulsion.
This effective repulsion takes the value of $U_{\rm eff} = 5.3$ $eV$ for the interaction parameters 
we have chosen for the HF calculation, which is very close to the critical value of 
$U_C / \left| \gamma_0 \right| = 2.23$ above which the choice would be unphysical
because the ground-state of a 2D graphene sheet develops an antiferromagnetic spin density wave solution \cite{fujita}.
For every choice in the cutoff for the Coulomb interaction range we could obtain
different effective choices of $U_{\rm eff}$ that can reproduce $\Delta^{AF} \left( k \right)$ satisfactorily.
In the Hubbard model the $k$ dependence of the diagonal terms of the
Hamiltonian in Eqs. (\ref{lhamil}, \ref{rhamil}) is manifestly due to the coefficients
$L_{k l}^{2}$ of the basis function with the largest contributions coming from the
lattice sites at the edge and decaying exponentially as we move into the bulk.
In the corrections due to farther neighbor interactions (i.e., beyond the Hubbard model) still the dominant
contribution to the matrix element $\Delta^{AF}$
for a given distance of electron interaction are those connecting
to the edge ribbon atom, resulting in a $k$-dependent behavior proportional
to that of the $L_{kl}^2$ located at the edge sites.
This remarkable behavior for which every additional neighbor contribution in the exchange potential 
has the same $k$-dependent behavior in the edge states region is most likely the reason for the good
overall agreement of the edge state properties in zigzag ribbons calculated within a simple Hubbard
model and other mean field calculations.
Outside this region of the Brillouin zone we find substantial differences of the potentials 
for the Hubbard model and Hartree-Fock hamiltonian matrix elements but these differences 
do not introduce relevant changes in the magnetic configuration of the edges.
From the results shown above we expect that the quantitative agreement of the gap at $ka \sim \pi$
between LDA approximation and the Hubbard calculations\cite{joaquin} with  $U = 2 eV$  suggests that 
the LDA will have a tendency to underestimate the energetics
of the ferromagnetic spin alignment along a zig-zag edge. In fact the LDA approximation \cite{perdew}
would introduce a larger gap opening if the spin density of the core electrons of carbon 
were considered when evaluating the spin dependent LDA exchange potential.

The off diagonal tunneling term $t_{\sigma} \left( k \right)$ given in equation
(\ref{tdef}) consists of a tight-binding term and an interaction term
coupling different sublattices. The onsite interaction term in the Hubbard model
cannot couple different sublattices, and therefore the two edge sites, and the matrix elements
reduces to the tight-binding band dispersion of the conduction edge band \cite{superexchange}. 
This matrix element can be enhanced in the presence of the long-ranged interaction.
The enhanced coherence between left and right edge solutions lead to an increase in the 
band-gap between the valence and conduction bands of the antiferromagnetic solution.
This enhancement of the tunneling term $t_{\sigma} \left( k \right)$ is more pronounced 
for $k$-points outside the edge states zone because the density tails spread more into the bulk.  
However it remains essentially zero in the surroundings of 
$ka \sim \pi$ since the wave functions are strongly localized at the ribbon edge borders.

The ribbon width dependence of both Hartree-Fock and Hubbard solutions remain similar.
Previously\cite{superexchange} we had illustrated using the Hubbard model that the
ribbon width dependence of the solutions can be summarized through the behavior
of the Hamiltonian matrix elements near the valley point $k \sim 2 \pi/3a$ that obey the
following scaling rules as a function of ribbon width $W$
\begin{eqnarray}
\Delta^{AF/F} \left( k \right) &=& W^{-1} \widetilde{\Delta}^{AF/F} \left( q W \right)  \\
 t \left( k \right) &=& \frac{\gamma_0}{W} \;  \tilde{t}\left( q W \right)
\end{eqnarray}
where $q = ka - 2\pi/3$ is the dimensionless momentum measured from the valley point
and the functions $\widetilde{\Delta}^{AF/F}$, $\widetilde{t}$ are functions that do not depend
on the width of the ribbon.
These rules for wide enough ribbons are not altered with the presence of
non-local exchange shown in equations (\ref{lhamil}, \ref{lrhamil}, \ref{rhamil})
because the additional non-local terms are quadratic in $L_{kl}$ and $R_{kl}$.
These in turn follow a similar scaling rule that can be obtained using a continuum model \cite{superexchange}
\begin{eqnarray}
R_{k \, l} &=&  W^{-1/2}\; \widetilde{R}_{qW,\,\, l} \\
L_{k \, l} &=&  W^{-1/2}\; \widetilde{L}_{qW, \,\, l}
\end{eqnarray}
where $\widetilde{R}_{qW \,\, l}$ and $\widetilde{L}_{qW \,\, l}$ are scaled functions that do not depend on the 
width of the ribbon.
Hence the band-gap size and the position of the gap as well as the up and down spin crossover point
of the F solution (measured from the valley point) all follow a $W^{-1}$ ribbon width dependence.

\section{Intra-edge localization properties of edge states}
We explore here the properties of localization of the zigzag edge states along
the $x$ axis, the direction of periodicity in the ribbon.
We estimate this quantity in real space evaluating a partial Fourier transfom of the $k$-point 
dependent density component centered at lattice site $l$ in the ribbon unit cell
\begin{equation}
F_{l \sigma} \left(x \right) = \int_{k > \left| 2\pi/3a  + 1/W \right|} dk \,\, e^{-i k x} \left| \Psi_{k l} \left( {\bf r} \right) \right|^2
\label{local}
\end{equation}
where $\Psi_{kl} \left( {\bf r} \right)$ is the $l^{th}$ component of edge band Bloch wave function.
The integration in $k$ space is over the region of edge states which give the relevant contribution
to the edge spin polarization.
This definition, reminiscent of the way Wannier functions are calculated, \cite{wannier}
shows us the way the electrons localize in real space along the edge.
\begin{figure}[htb]
\begin{center}
\begin{tabular}{c}
\resizebox{80mm}{!}{\includegraphics{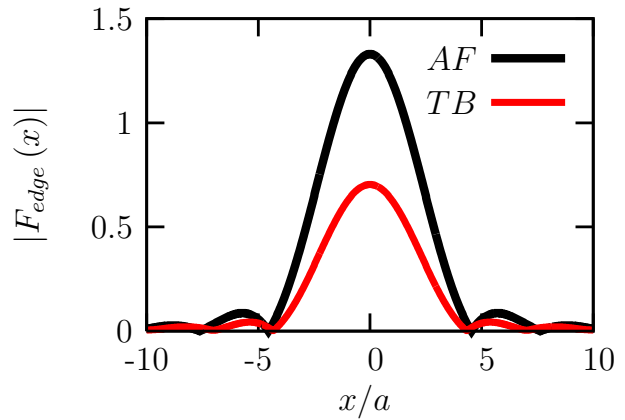}}  
\end{tabular}
\caption{
(Color online)
Wave function localization along the ribbon centered at
different sites in the ribbon unit cell calculated for a ribbon with $N = 12$
corresponding to the antiferromagnetic ground state solution and interaction 
free tight-binding solution.
We can notice a substantial spreading of the wave function at the ribbon
edge enveloping about 7 units of lattice constant $a$ along the ribbon.
}
\label{kpoint2}
\end{center}
\end{figure}
In figure (\ref{kpoint2}) we focus on the row of atoms at the ribbon edge and we show that the 
edge state wave functions spread about seven atomic lattice sites on average, and this 
spreading length is essentially same for paramagnetic tight-binding solutions and
the solutions with edge spin polarization. 
These features are robust to changes in the ribbon width and the
details of the Coulomb interaction with negligibly small departures from this form
when we change the strength of the Coulomb interaction
or modify the range of the Coulomb interaction.
This robust feature of the wide smearing of the electron density on the
edge atoms into the neighboring atoms located several lattice constants away along
the edge direction is key in producing the long reaching ferromagnetic spin correlation length
in a zigzag ribbon edge. \cite{fujita,yazyev}
Due to this far reaching overlap of the electron density
along the edge even short ranged exchange interactions are able to connect
edge states centered around different edge atoms and induce an energetically 
favorable parallel spin alignment.

An intuitive way to relate this edge wave function localization length along the edge
with the band structure or the ribbons near $k a \sim \pi$ is
by relating the gain of exchange energy per particle for the occupied
electrons near the edge atoms through $\Delta_{max} = \Delta^{AF/F} \left( \pi / a \right) $
with an effective localization length $\lambda$ through the formula
\begin{equation}
{\Delta_{max}}  = \frac{e^2}{ \epsilon_r \lambda } .
\end{equation}
We present in table \ref{lambda_table} the values of $\Delta_{max}$
and $\lambda$ for different choices of $\epsilon_r$ calculated for
a ribbon of $N = 12$ atom pairs width. In this analysis we define the onsite term
as $U = 10 / \epsilon_r \,\,\,\, eV$ making it dependent of the dielectric screening. 
The value of $\Delta_{max}$ remain practically constant as we make the ribbon wider and
already for $N = 12$ it gives a good estimate of the infinite width limit.
\begin{table}[ht]
\centering 
\begin{tabular}{l| c c c c c}
\hline\hline                        
$\epsilon_r$  &  1 & 2 & 3 & 4 & 5 \\ [0.5ex]
\hline 
$\Delta_{max}$  & 2.35   & 1.16 & 0.78 & 0.58 & 0.46  \\
$\lambda$       &  2.49  &  2.52 & 2.5  & 2.5  &  2.5 \\   [1ex] 
\hline 
\end{tabular}
\caption{Effective localization length $\lambda$ (in units of lattice constant $a = 2.46 \AA$) 
estimated from the shift from tight-binding bands $\Delta_{max}$ (in $eV$) near $k a \sim \pi$ 
for antiferromagnetic spin configuration in graphene nanoribbons.
We can observe that $\lambda$ does not change much as a function of the
values of dielectric constants considered.
We can observe that the magnitudes of $\lambda$ are consistent with the
estimations that can be extracted for the localization lengths presented in
the previous section.}
\label{lambda_table}
\end{table}
The resulting values of $\lambda$ shows only a very small variation when
different values of the relative dielectric constant $\epsilon_r$ are used,
in agreement with the fact that the shapes of the density localization
remains practically unchanged.


\section{Energetics of inter-edge exchange coupling}    
The nature of antiferromagnetic spin polarization of edge states on different sublattices 
(and therefore different edges) was studied previously as an unusual type of 
superexchange interaction \cite{superexchange}
where it is energetically favored with respect to the ferromagnetic state 
both in the kinetic energy and interaction energy.
For the Hubbard model the inter-edge coupling is mediated by the band Hamiltonian 
tunneling and the local spin-polarization is provided by intra-edge exchange.
In the presence of non-local interactions 
the inter-edge off-diagonal tunneling term plays a role in determining the total exchange energy.
In this section we examine the impact of non-locality of exchange in the energetics of the
ground-state solution by obtaining the differences in the total energies between
AF and F mean field solutions.
In a neutral nanoribbon the electrostatic Hartree energy is the same in both AF and F configurations
and the total energy difference per edge carbon atom $\Delta E$ consists of the kinetic energy
term and the exchange term
\begin{eqnarray}
\Delta E = E^{F} - E^{AF} = \Delta T + \Delta E_{X}.
\end{eqnarray}
The kinetic energy difference can be calculated from
one-body averages of the tight-binding Hamiltonian
for each occupied edge-band states and is essentially the same derivation
as in reference [\onlinecite{superexchange}] while for the exchange energy we need to sum two body exchange integrals.
It will be useful to write it as an integral of a $k$-dependent function
defining implicitly $\epsilon_{X} \left( k \right)$
\begin{eqnarray}
E_{X} = - \frac{1}{2} \sum_{i,j}^{occ} K_{ij}
= \int_{BZ} dk \,\, \epsilon_{X} \left( k \right)
\label{xen}
\end{eqnarray}
where $K_{ij}$ is the standard definition of exchange integral as can be found in reference [\onlinecite{ostlund}].
The labels $i$, $j$ represent occupied single particle states
including the $k$ quantum numbers, band labels and spin index. 
In Fig. (4) we show the exchange energy difference as a
function of $k$-point in the Brillouin zone
for the Hubbard model and the Hartree-Fock approximation.
\begin{figure}[htbp]
\label{ex_distr}
\begin{center}
\begin{tabular}{cc}
\includegraphics[width=4.45cm]{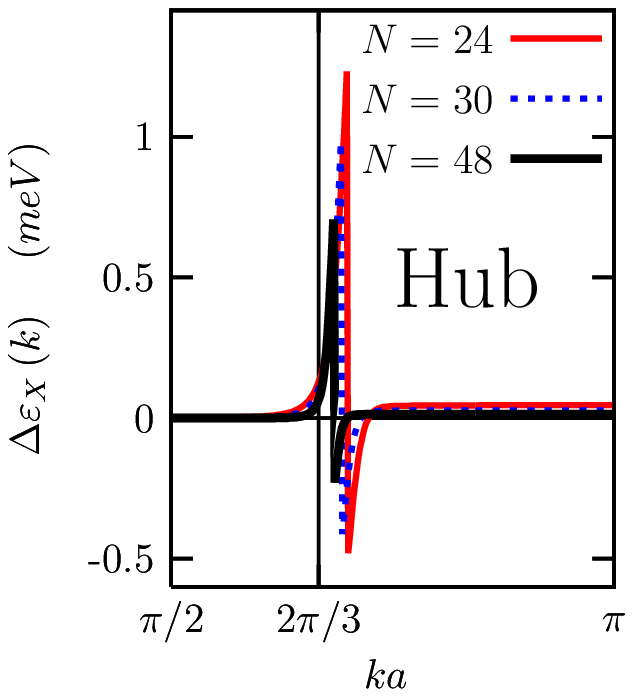} $\quad$&
\includegraphics[width=3.55cm]{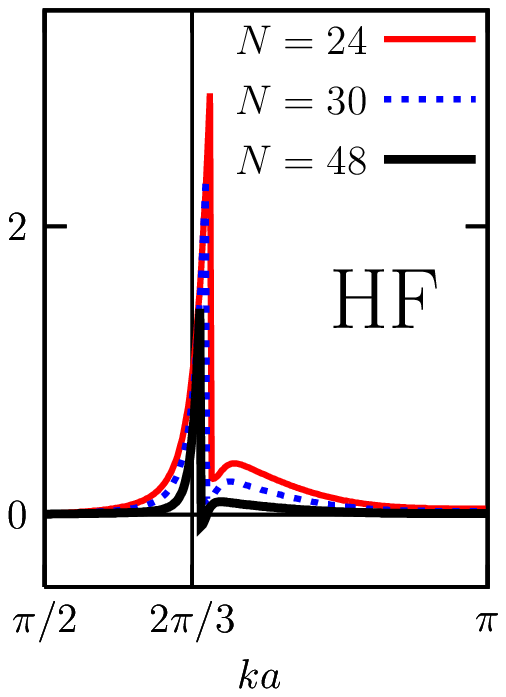}
\end{tabular}
\caption{
(Color online)
Comparison of $k$-point resolved exchange energy differences
between Hubbard and HF calculations.
{\em Left panel:}
Exchange energy difference density
$\Delta \epsilon_{X} \left( k \right)$ for Hubbard calculation with $U = 2 eV$
Most of the exchange energy difference comes from the regions of
$ \left| k \right|  \sim 2\pi/3a$ while the region $ka \sim \pi$ also adds
a small contribution. 
{\em Right panel:}
In presence of longer ranged interactions we find contributions to the energy
differences in a wider range of $k$-points around the valley point $2 \pi /3a$
thanks to the enhanced intra-edge tunneling terms.
}
\label{figex}
\end{center}
\end{figure}
We can observe that the details of $k$-point dependent contributions to the exchange 
energy difference is substantially modified in presence of non-local interactions.
The global enhancement in magnitude leads to a further stability of AF solutions.
In particular the off diagonal interaction terms in the effective Hamiltonian further enhances 
the stability of AF solutions through an enhanced inter-edge tunneling.

The ribbon width dependence of the total energy differences can be derived
from similar considerations as in reference [\onlinecite{superexchange}] and
follows a $W^{-2}$ decay law.
\begin{figure}[htbp]
\begin{center}
\begin{tabular}{c}
\includegraphics[width=8cm]{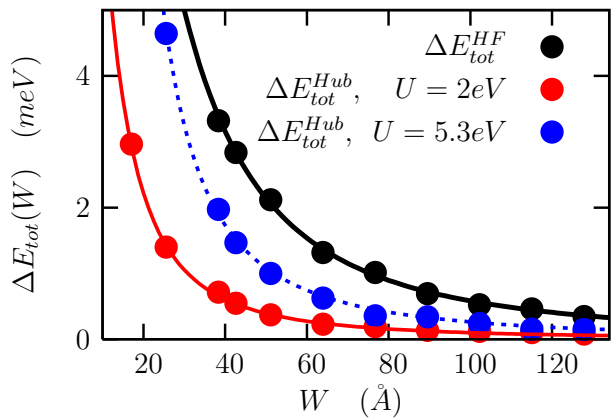}
\end{tabular}
\caption{
(Color online)
Total energy difference per edge carbon atom as a function of ribbon width showing a
$W^{-2}$ decay behavior.
The fitting curves in units of $eV$ versus $\AA$ are
$6/(W^2 + 300)$ for $\Delta E^{HF}_{tot}$,
$1/(W^2 + 50)$ for $\Delta E^{Hub}_{tot}$ with $U = 2eV$
and  $2.6/(W^{2} - 100)$ for $U = 5.3eV$.
}
\label{interedge}
\end{center}
\end{figure}

\section{Summary and conclusions}
In this paper we have addressed the effects of non-locality of exchange in
the edge states of a neutral zig-zag graphene ribbon from different points of view.
We started with a formal introduction on how the far reaching interaction terms between
distant sites can influence the values of the matrix elements of the effective two-bands Hamiltonian
describing the edge states, distinguishing the different roles in terms of
intra-edge and inter-edge interaction manifested respectively in the diagonal and off
diagonal matrix elements, representing respectively the direct exchange energy
giving rise to the spin polarization and the tunneling between both sublattices
that mix states localized at different edges.
Further insight was gained studying the properties of localization of edge states
along the direction of the ribbon finding for those states sitting on the edge atoms
a rather long ranged enveloping function in the direction of periodicity, a fact that would
make possible even for very short ranged interaction terms to connect with the wave functions
centered several lattice constants away in the direction of the ribbon.
Finally we analyzed the impact of non-local interaction in the
energetics of inter-edge coupling largely responsible for the antiferromagnetic
spin polarization of the system.

In light of our studies we have been able to understand better why the Hubbard
model based on a strictly short ranged onsite interaction has turned out to give
results consistent to solutions found in more elaborate studies.
The main reasons would be on the one hand that for edge states the inter-edge tunneling
not captured by the Hubbard model is relatively less important with respect to
the intra-edge spin splitting contributions whose functional form in $k$-space is
accurately captured by an onsite interaction.
On the other hand, the spreading of the edge state wave functions along the ribbon allows 
edge states centered at different atomic sites to connect each other even for strictly short ranged onsite interactions. 
Edge width depending scaling of band gaps and energy differences of different spin polarized states
do also follow a similar law of $W^{-1}$ and $W^{-2}$ respectively as found in the Hubbard model
since the width dependence of those quantities are dictated by the
wide ribbon asymptotic behavior of the edge state wave functions. \cite{superexchange}
Even though the above mentioned qualitative features remain the same both for non-local
and local exchange described by the HF and Hubbard models
we have also shown that the discrepancies in the total energy differences 
and the exchange potentials outside the edge states region of the Brillouin zone
cannot be modeled accurately with a renormalized effective onsite interaction term. 
This effect can have a relevant influence in calculating the the spin stiffness \cite{yazyev, rhim} 
or in the band structure of the system when it is shifted away from the neutrality point.\cite{doping,magnetoelectric} 
The analysis we carried out illustrates the role of non-local exchange in altering the 
band dispersion and band-gap size due to spin polarized edge states,
explaining the relative enhancement of band gaps in the results obtained with 
non-local B3LYP \cite{pisani} type functionals with respect to those obtained 
by more local prescriptions like LDA/GGA functionals. \cite{son_gap}

\section{Acknowledgment. }
The author gratefully acknowledges valuable discussions with T. Pereg-Barnea and A. H. MacDonald.
Financial support was received from Welch Foundation grant TBF1473, 
NRI-SWAN, DOE grant Division of Materials Sciences and Engineering DE-FG03-02ER45958.

\newpage
\section*{Appendix I. Hartree-Fock formulation in the two band analysis}
The full Hartree-Fock Hamiltonian can be reduced to a two dimensional matrix representation
for each spin. We will
represent this Hamiltonian matrix in a suitably chosen basis function, for example
the LR basis we have defined in the main text.
Given the basis of Bloch functions
\begin{equation}
\label{basis}
 \left< {\bf r} | k \lambda \right> = \psi_{{\bf k} \lambda} \left({\mathbf r} \right)
= \frac{1}{\sqrt{N_K}} \sum_{i} e^{i {\mathbf k} \left( {\mathbf R}_i + {\bf \tau}_{l} \right)}
\phi \left({\mathbf r} - {\mathbf R}_i - {\bf \tau}_{l} \right) \eta_{\sigma}
\end{equation}
where $\eta_{\sigma}$ represents the spinor, ${\bf \tau}_{l}$ represents the
displacement vector for sublattices in the unit cell that can be labeled with $l$,
and $N_{K}$ is the total number of $k$-points, or equivalently, the number of
unit cells in the system repeated in the periodic direction.
The label $\lambda = \left( l, \sigma \right)$ represents both the lattice
site label $l$ and the spin $\sigma$.
In this basis the general expression of the Hamiltonian is
\begin{eqnarray}
\label{hfgen}
V_{HF} &=& \sum_{k \lambda \lambda^{\prime}} U_H^{\lambda \lambda^{\prime}}
\left[ \sum_{k^{\prime}}
\left<  c^{\dag}_{k^{\prime} \lambda^{\prime}} c_{k^{\prime} \lambda^{\prime}} \right>  \right]
c^{\dag}_{k \lambda} c_{k \lambda}     \nonumber
\\
&-& \sum_{k^{\prime}\lambda \lambda^{\prime}} U_{X}^{\lambda \lambda'}
\left(k^{\prime} - k \right)
\left<  c^{\dag}_{k^{\prime} \lambda^{\prime}} c_{k^{\prime} \lambda} \right>
c^{\dag}_{k \lambda} c_{k \lambda^{\prime}}
\end{eqnarray}
where
\begin{eqnarray}
U_H^{\lambda \lambda'} &=&
\left< {\bf k} \lambda {\bf k}' \lambda' \left| V \right| {\bf k} \lambda {\bf k}' \lambda' \right>  \\
&=& \int d r_1 d r_2 \left| \psi_{k \lambda} \left( r_1 \right) \right|^2  V\left(r_1, r_2 \right)
\left| \psi_{k' \lambda'} \left( r_2 \right) \right|^2  \left( r_2 \right)    \nonumber
\end{eqnarray}
and
\begin{eqnarray}
U_X^{\lambda \lambda'} \left( {\bf q} \right) &=&
\left< {\bf k} \lambda {\bf k}' \lambda' \left| V \right| {\bf k}' \lambda {\bf k} \lambda' \right>  \\
&=& \int d r_1  d r_2 \psi_{k \lambda}^{*} \left( r_1 \right) \psi_{k^{\prime} \lambda}
\left( r_1 \right) \\
&\times& V\left(r_1, r_2 \right)
\psi^{*}_{k^{\prime} \lambda} \left( r_2 \right) \psi_{k \lambda^{\prime}} \left( r_2 \right)    \nonumber
\end{eqnarray}
where $\psi_{k \lambda} \left( r \right)$ are the Bloch state wave functions.

The matrix elements in the Left-Right (LR) basis can be written in the following way
\begin{equation}
\label{hartree}
\left< k L \right| V_H \left| k L \right> = \sum_{\lambda} \left| L_{k \lambda}\right|^2
 \sum_{ k^{\prime} \lambda^{\prime}}
N_{K} U^{\lambda \lambda^{\prime}}_H \left<  n_{k^{\prime} \lambda^{\prime} \lambda^{\prime}} \right> 
\end{equation}
where $N_{K}$ is the total number of $k$ points used in the sum.
The density matrix operator is defined as
$n_{\lambda \lambda^{\prime}} = c^{\dag}_{\lambda^{\prime}} c_{\lambda}$
where the label $\lambda$ represents the quantum numbers that labels the basis set.
The explicit forms of the amplitude coefficients $L_{k l}$ in the tight-binding
model can be found in the appendix II.
The diagonal $n_{\lambda} \equiv n_{\lambda \lambda}$ is simply the occupation
in the state $\lambda$ and implicitly implies a sum in the $k$ points
\begin{equation}
\label{denmat}
n_{\lambda \lambda^{\prime}}
= \frac{1}{N_{K}} \sum_{k} c^{\dag}_{k \lambda} c_{k \lambda}.
\end{equation}

We get a similar expression for the other element of the diagonal where we only
need to change the $L$ label into $R$.
The Hartree term averages to zero for the off diagonal matrix element.
In a neutral ribbon the presence of the positive background charge neutralizes the
electrostatic Hartree-potential
\begin{eqnarray}
\widetilde{V}_{ext}\left( l \right) + \widetilde{V}_{H} \left( l \right) = 0
\end{eqnarray}
so we can focus our attention in the exchange
contribution of the interaction.

Now we show the matrix elements of the Fock term.
For the diagonal and off diagonal elements we have respectively
\begin{widetext}
\begin{eqnarray}
\label{fock}
\left< k L \right| V_F \left| k L \right> &=&
\left<  k L  \right|
- \sum_{k^{\prime} \lambda \lambda^{\prime}} U_{X}^{ \lambda \lambda'} \left( k^{\prime} - k \right)
\left< c^{\dag}_{k^{\prime} \lambda^{\prime} } c_{k^{\prime} \lambda } \right> c^{\dag}_{k \lambda} c_{k \lambda^{\prime}}
\left| k L \right>  
 = - \sum_{\lambda \lambda^{\prime}} L_{k \lambda}^* L_{k \lambda^{\prime}}
\sum_{k^{\prime}} U_X^{ \lambda \lambda'} \left( k^{\prime} - k \right)
\left< n_{k^{\prime} \lambda \lambda^{\prime}} \right>
\end{eqnarray}
\begin{eqnarray}
\label{fock1}
\left< k L \right| V_F \left| k R \right> &=&
\left<  k L  \right|
- \sum_{k^{\prime} \lambda \lambda^{\prime}} U_{X}^{\lambda \lambda^{\prime}} \left( k^{\prime} - k \right)
\left< c^{\dag}_{k^{\prime} \lambda^{\prime} } c_{k^{\prime} \lambda } \right> c^{\dag}_{k \lambda} c_{k \lambda^{\prime}}
\left| k R \right>    
 = - \sum_{\lambda \lambda^{\prime}} L_{k \lambda}^* R_{k \lambda^{\prime}}
\sum_{k^{\prime}} U_X^{\lambda \lambda^{\prime}} \left( k^{\prime} - k \right)
\left< n_{k^{\prime} \lambda \lambda^{\prime}} \right>
\end{eqnarray}
\end{widetext}
differing only in the expansion coefficients of the two-bands states.

The kernels in the Hartree and Fock terms of equation (\ref{hfgen}), 
assuming $\sigma = \sigma^{\prime}$ and dropping the spin part,
can be written as 
\begin{widetext}
\begin{eqnarray}
U_H^{\lambda \lambda'} &=& \frac{1}{N_K^2} \sum_{i,j}^{N_K}
\int d{\bf r}_1 d{\bf r}_2   
 \left| \phi \left({\bf r}_1 - {\bf R}_i - \tau_{l}  \right) \right|^2
 v\left( {\bf r}_1 - {\bf r}_2 \right)
 \left| \phi \left({\bf r}_2 - {\bf R}_j - \tau_{l'}  \right) \right|^2     
\simeq
\frac{1}{N_K^2}
\sum_{i,j}^{N_K}  V_{\rm eff}\left(  \left| {\bf L}^{l l^{\prime}}_{ij} \right| \right) \\
U_X^{\lambda \lambda'} \left( {\bf q} \right) &=&
\frac{1}{N_K^2}  \sum_{i^{\prime} j^{\prime}}^{N_K}
e^{i  \left(  k^{\prime} - k \right)    \left( R_{i^{\prime}} + \tau_{l} - \left( R_{j^{\prime}} + \tau_{l'} \right) \right) }
V_{\rm eff} \left( \left| R_{i^{\prime}} + \tau_l  - R_{j^{\prime}} - \tau_{l'}  \right| \right)    
\simeq
\frac{1}{N_K^2} \sum_{ij}^{N_K}
e^{i \left({\bf k}' - {\bf k} \right) {\bf L}_{ij}^{l l^{\prime}} }
\,\,\, V_{\rm eff}\left( \left| {\bf L}_{ij}^{ll'} \right| \right)
\end{eqnarray}
\end{widetext}
where we have defined
${\bf L}^{l l^{\prime}}_{ij} =  {\bf R}_i - {\bf R}_j + \tau_{l} - \tau_{l^{\prime}}$
and the Coulomb integrals $V_{\rm eff}\left( \left| {\bf L}_{ij}^{ll'} \right| \right)$ can be approximated through the expression
in equation (\ref{coulint}).

The Hartree-Fock equations reduce to the Hubbard model when the interactions 
are reduced to onsite repulsion and the expressions simplify considerably.
\begin{eqnarray}
\left< k L \right| V_H \left| k L \right> &=& \sum_{l} \left| L_{kl}\right|^2
U \left(  \left< n_{l \sigma}\right>  + \left< n_{l \overline{\sigma}}\right>   \right)  \\
\left< k L \right| V_F \left| k L \right> &=&
- U \sum_{l} \left| L_{kl} \right|^2 \left< n_{l \sigma} \right>.
\end{eqnarray}
Adding both terms we obtain for the diagonal and off diagonal terms
\begin{eqnarray}
\left< k L \right| V_{HF} \left| k L \right>  &=& U \sum_l \left| L_{kl} \right|^2 \left< n_{l \overline{\sigma}} \right> \\
\left< L \right| V_F \left| R \right> &=& - U  \sum_{l } L_{kl}^* R_{k l} \left< n_{ l \sigma} \right>.
\end{eqnarray}
The last off diagonal term reduces to zero because the coefficients $L_{kl}$ and $R_{kl}$ have
zero overlap.

\section*{Appendix II. Analytic resolution of the tight-binding LR-functions}
In this appendix section we describe the exact tight binding wave functions for the zigzag ribbon edge states.
We use periodic boundary conditions in the x-direction and closed boundary conditions in the y direction.  Taking advantage of the translational symmetry we may Fourier transform the Hamiltonian and solve a one dimensional problem with $k\equiv k_x$ as a parameter.  This derivation is similar to that of Malysheva {\it et al.}\cite{malysheva} with some additional details.

The $k$-dependent one dimensional Hamiltonian of graphene is given by:
\begin{eqnarray}
{\cal H}_{k}&=& \Psi^\dagger_{k} H_{k} \Psi_{k} \\
H_{k} &=& \begin{pmatrix}
0 & R + Q \\ R^\dagger + Q & 0
\end{pmatrix}
\end{eqnarray}
where $Q = 2\cos(k/2)$ and the vectors $\Psi$ have $2N$ coordinate, the $N$ top coordinates are the A sublattice sites along the y-direction and the bottom $N$ are the B sublattice sites.  The operator $R^\dagger$ is a $N\times N$ matrix which represents a translation by one unit cell in the positive y-direction and $R$ translates in the opposite direction.  Employing the ansatz:
\begin{equation}
\Psi_{k}(n) = \begin{pmatrix}
\psi_A \\ \psi_B
\end{pmatrix} z^n
\end{equation}
we may reduce Schr\"{o}dinger's equation to:
\begin{eqnarray}
(1/z+Q)\psi_B = E\psi_A \nonumber \\
(z + Q)\psi_A = E\psi_B
\end{eqnarray}
In order to solve these equations with a non-trivial vector $(\psi_A,\psi_B)$ we require that the matrix of coefficients have a zero determinant.  This leads to a relation between $z$ and the energy $E$:
\begin{eqnarray}\label{eq:energy}
\left|\begin{pmatrix} -E & z+2\cos(k/2) \\
 1/z + 2\cos(k/2)& -E \end{pmatrix}\right| =0 \nonumber \\
 \nonumber  \\
E^2 = 1+(z+1/z)Q + Q^2
\end{eqnarray}
Note that for a solution $z$, $1/z$ is also a solution.  We can therefore, write two (unnormalized) solutions:
\begin{eqnarray}\label{eq:solutions}
\Psi_1(z,E,k) = \begin{pmatrix}  \frac{1}{ z} + Q \\ E \end{pmatrix} z^n\\ \nonumber
\Psi_2(z,E,k) = \begin{pmatrix}  z+ Q \\ E \end{pmatrix} z^{-n}
\end{eqnarray}

In order to complete our description of wave functions we need one more condition on $E$ or $z$.  This condition comes from the edges of the ribbon.  For pedagogical purposes, let us briefly mention the case of infinite and semi-infinite samples.  In an infinite system we require that the wave functions stay finite in the limits $n\to\pm\infty$.  This leads to $|z|=1$ which is obeyed by Bloch waves $z^n = e^{ipn}$ with a real $p$ being the momentum along the $y$-direction.  In a semi-infinite sheet we require only that the wave function be finite at $n\to +\infty$ and a vanishing amplitude on the one edge.  In order to satisfy the vanishing amplitude condition we need to combine the two solutions in Eq.(~\ref{eq:solutions}) and in order to satisfy the condition at infinity we require $|z|\leq 1$.  Bloch waves (with $|z|=1$) can satisfy both conditions thus the bulk states are rearranged to accommodate the edge.  In addition, new real solutions with $|z| < 1$ may also satisfy the boundary condition on the edge with $E=0$.  This gives either $z = -2\cos(k/2)$ or $1/z = -2\cos(k/2)$ when only one of these solutions is physical and decays away from the edge.  The vanishing amplitude condition is trivially satisfied by setting all A (or B) site amplitudes to zero.  This gives the flat band states at momenta $k$ which satisfy $|\cos(k/2)|<1/2$.  This condition defines the region between the $K$ and $K'$ points.

In the case of a ribbon with two edges we require a vanishing amplitude on both sides of the ribbon.  This translates to $\Psi_{k}^B(0) = \Psi_{k}^A(N+1) = 0$ where $N$ is the number of unit cells in the $y$ direction (i.e., the number of chain pairs, as defined in the main text).  Assuming that $\Psi_k(n)$ is a linear combination of the solutions in Eq.(\ref{eq:solutions}) we obtain the relation:
\begin{eqnarray}\label{eq:z}
-Q = \frac{z^{N}-z^{-N}} { z^{N+1}-z^{-(N+1)}}.
\end{eqnarray}
which determines all the possible solutions of the tight binding model on the ribbon.  Eq.(\ref{eq:z}) has $N-2$ Bloch like solutions and $2$ real (edge like) solutions\cite{malysheva}. For real numbers $z^n$ produce exponentially decaying solutions which are localized to one of the edges.  Similarly to the case of the semi-infinite graphene sheet, the real, edge-localized solutions are found in a region between the valleys.  The term valleys usually refers to the nodes in the Brillouin zone where the energy vanishes.  Here, we can only see their projection on the $k$
axis at $k = \pm \frac{2\pi}{3a}$.  However, due to the finite size of the ribbon the edge states regime is smaller and extends from
$2 \pi/3a + 1/W$ to $4 \pi/3a - 1/W$ as will be shown shortly.

In order to find the amplitudes of the left and right wave functions all we need to do is combine the two solutions in Eq.(\ref{eq:solutions}) in a way that the boundary conditions are satisfied.  We may choose to satisfy the boundary condition provided by one of the edges while the other condition would be automatically satisfied by the correct choice of $z$ (a solution to Eq.(\ref{eq:z})).
This gives:
\begin{eqnarray}
\Psi_\pm(E,k,n) = \alpha \left( \begin{pmatrix} \frac{1}{z}+Q \\ \pm |E| \end{pmatrix}
z^n-
\begin{pmatrix} z+Q \\ \pm|E| \end{pmatrix}z^{-n} \right)
\end{eqnarray}
and the parameter $\alpha $ is given by the normalization:
\begin{eqnarray}
|\alpha|^{-2} &=&  \sum_{n=1}^N [ (( \frac{1}{z}+Q)z^n-(z+Q)z^{-n})^2 + E^2(z^n-z^{-n})^2]\nonumber \\
&=& 2(1-z^2)  \frac{(1-z^{2N})(1+z^{2(N+1)})}{(1-z^{2(N+1)})^2}
\end{eqnarray}
Please note that $z$ is assumed real (but may be negative).  It is worth mentioning that the lower energy solution (with $-|E|$) is antisymmetric about the ribbon center while the higher energy solution is symmetric.  For this (anti)symmetry to hold we require $\Psi^A(n) = \pm\Psi^B(N+1-n)$ which is satisfied by $\Psi_\pm$.

To gain some intuition about interacting edge states we prefer to work with edge localized states $\Psi_L$ and $\Psi_R$. These are constructed by combining the $\Psi_\pm$.  Since the parameters $z,Q$ and $ \alpha$ only depend on the absolute value of $E$ we get:
\begin{eqnarray}
\Psi_L &=& \frac{1}{\sqrt{2}} (\Psi_++\Psi_-) = \sqrt{2}\alpha(( \frac{1}{z} +Q)z^n-(z+Q)z^{-n})
\begin{pmatrix} 1 \\ 0 \end{pmatrix} \nonumber \\
\Psi_R &=& \frac{ 1}{\sqrt{2}} (\Psi_+-\Psi_-)= \sqrt{2}|E|\alpha(z^n-z^{-n})
\begin{pmatrix} 0 \\ 1 \end{pmatrix}
\end{eqnarray}
Note that the left and right wave functions are localized on one sublattice, determined by the boundary conditions.

In the above discussion we left $z$ as a parameter.  However, the value of $z$ is defined by the solutions to Eq.\ref{eq:z}, which unfortunately does not provide a closed form. Let us define the inverse localization length, $\lambda$ such that $z = \exp(\lambda)$.  Eq.\ref{eq:z} can be rewritten as
\begin{equation}\label{eq:lambda}
-Q = {\sinh(N\lambda) \over \sinh((N+1)\lambda)}.
\end{equation}
Then the energy of this localized solution can be simplified by substituting Eq.(\ref{eq:lambda}) into Eq.(\ref{eq:energy}):
\begin{equation}\label{eq:Esinh}
E = \pm {\sinh(\lambda) \over \sinh((N+1)\lambda)}.
\end{equation}


\begin{thebibliography}{99}
\bibitem{fujita} 
                M. Fujita, K. Wakabayashi, K. Nakada, K. Kusakabe,
                J. Phys. Soc. Jpn. {\bf 65}, 1920 (1996).

\bibitem{nakada} 
               K. Nakada, M. Fujita, G. Dresselhaus, M. S. Dresselhaus,
               Phys. Rev. B {\bf 54}, 17954 (1996).

\bibitem{waka} 
             K. Wakabayashi, M. Fujita, H. Ajiki, M. Sigrist,
             Phys. Rev. B {\bf 59}, 8271 (1999).

\bibitem{ezawa} M. Ezawa,
              Phys. Rev. B {\bf 73}, 045432 (2006).

\bibitem{brey}L. Brey and H. A. Fertig,
              Phys. Rev. B {\bf 73}, 235411 (2006).


 \bibitem{hikihara}  T. Hikihara et al.,
Phys. Rev. B {\bf 68}, 035432 (2003).

\bibitem{son_gap} Y.-W. Son, Marvin L. Cohen, and Steven G. Louie,
               Phys. Rev. Lett. 97, 216803 (2006).

\bibitem{son_half} Y.-W. Son, Marvin L. Cohen, and Steven G. Louie,
               Nature {\bf 444}, 347 (2006).

\bibitem{pisani} L. Pisani, J. A. Chan, B. Montanari, and N. M. Harrison
Phys. Rev. B {\bf 75}, 064418 (2007).

\bibitem{leehosik} H. Lee, Y.-W. Son, N. Park, S. Han, and J. Yu,
Phys. Rev. B {\bf 72}, 174431 (2005).

\bibitem{kusakabe}
K. Kusakabe and M. Maruyama,
Phys. Rev. B 67, 092406 (2003).  

\bibitem{yazyev} O. Yazyev, M. I. Katsnelson
Phys. Rev. Lett. {\bf 100}, 047209 (2008).

\bibitem{joaquin} J. Fern\'{a}ndez-Rossier,
Phys. Rev. B {\bf 77}, 075430 (2008).

\bibitem{pohang}
W. Y. Kim and  K. S. Kim,
Nature Nanotechnology {\bf 3}, 408  (2008). 

\bibitem{superexchange}
J. Jung. T. Pereg-Barnea, A. H. MacDonald, Phys. Rev. Lett. {\bf 102}, 227205 (2009). 

\bibitem{rhim}
J.-W. Rhim and K. Moon,
Phys. Rev. B {\bf 80}, 155441 (2009).


\bibitem{white}
D. Gunlycke, D. A. Areshkin, L. Junwen, J. W. Mintmire, C. T. White,
Nano Lett. {\bf 7}, 3608 (2007);  
D. Gunlycke, H. M. Lawler, C. T. White, 
Phys. Rev. B. 75, 29-33 (2007).

\bibitem{sandler}
M. Zarea, C. Busser and N. Sandler, Phys. Rev. Lett. 101, 196804 (2008).

\bibitem{sudipta}
S. Dutta, S. Lakshmi and S. K. Pati,
Phys. Rev. B 77, 073412 (2008).


\bibitem{doping}
J. Jung and A. H. MacDonald, Phys. Rev. B 79, 235433 (2009).


\bibitem{magnetoelectric}
J. Jung and A. H. MacDonald, Phys. Rev. B 81, 195408 (2010).


\bibitem{sudipta1}
Sudipta Dutta and Swapan K. Pati,
J. Mater. Chem., {\bf 20}, 8207 (2010).




\bibitem{xu}
B. Xu, J. Yin, H. Weng, Y. Xia, X. Wan, and Z. Liu,  
Phys. Rev. B 81, 205419 (2010).

\bibitem{qin}
R. Qin, J. Lu, L. Lai, J. Zhou, H. Li, Q. Liu, G. Luo, L. Zhao, Z. Gao, W. N. Mei, and G. Li,
Phys. Rev. B 81, 233403 (2010).  

\bibitem{kuntsmann}
J. Kuntsmann, C. Ozdogan, A. Quandt, and H. Feshke,
Phys. Rev. B {\bf 83}, 045414 (2011).


\bibitem{niimi}
Y. Niimi, T. Matsui, H. Kambara, K. Tagami, M. Tsukada, and H. Fukuyama,
Phys. Rev. B {\bf 73}, 085421 (2006).



\bibitem{kobayashi}
Y. Kobayashi, K.-I. Fukui, and T. Enoki, and K. Kusakabe,
Phys. Rev. B {\bf 73}, 125415 (2006).


\bibitem{han}
M. Y. Han, Barbaros \"Ozyilmaz, Y. Zhang, and P. Kim, Phys. Rev. Lett. 98, 206805 (2007).

\bibitem{chem_ribbon} 
X. Li, X. Wang, L. Zhang, S. Lee, H. Dai,
Science {\bf 319}, 1229 (2008).

\bibitem{etching}
S. S. Datta, D. R. Strachan, S. M. Khamis and A. T. C. Johnson,
Nano Lett. {\bf 8} 1912 (2008).

\bibitem{jarillo}
L. C. Campos, V. R. Manfrinato, J. D. Sanchez-Yamagishi, J. Kong, P. Jarillo-Herrero,
Nano letters. 9, 2600 (2009).

\bibitem{rice}
L. Ci, L. Song, D. Jariwala, et al. 
Adv. Mat. 21, 4487 (2009).


\bibitem{dresselhaus}
X. Jia, M. Hofmann, V. Meunier, B. G. Sumpter, J. Campos-Delgado, J. M. Romo-Herrera, H. Son, Y.-P. Hsieh, A. Reina, J. Kong, 
M. Terrones, M. S. Dresselhaus,
Science {\bf 323}, 1701 (2009).


\bibitem{zettl}
C. \"O. Girit, J, C. Meyer, R. Erni, M. D. Rossell, C. Kisielowski, L. Yang, C.-H. Park, M. F. Crommie, M. L. Cohen, S. G. Louie, and A. Zettl,
Science {\bf 323}, 5922 (2009).


\bibitem{chuvilin}
A. Chuvilin, J. C. Meyer, G. Algara-Siller, U. Kaiser, 
New Journal of Physics 11, 083019 (2009).

\bibitem{ye}
M. Ye, Y. T. Cui, Y. Nishimura, Y. Yamada, S. Qiao, A. Kimura, M. Nakatake, H. Namatame and M. Taniguchi,
Eur. Phys. J. B {\bf 75}, 31 (2010).

\bibitem{enoki}
T. Enoki, and K. Takai, Solid State Commun. {\bf 149}, 1144  (2009).




\bibitem{joly}
V. L. J. Joly, M. Kiguchi, S.-J. Hao, K. Takai, T. Enoki,
R. Sumii, K. Amemiya,
H. Muramatsu, T. Hayashi, Y. A. Kim, and M. Endo,
J. Campos-Delgado, F. L—pez-Ur'as,
A. Botello-MŽndez, H. Terrones,          
M. Terrones,
M. S. Dresselhaus,
Phys. Rev. B {\bf 81}, 245428 (2010).

\bibitem{crommie}
C. Tao, L. Jiao, O. V. Yazyev, Y.-C. Chen, J. Feng, X. Zhang, R. B. Capaz, J. M. Tour, A. Zettl, S. G. Louie, H. Dai, M. F. Crommie
arXiv:1101.1141v1 (2010).




\bibitem{novoselov}
%
K. S. Novoselov, A. K. Geim, S. V. Morozov, D. Jiang, M. I. Katsnelson, I. V. Grigorieva, S. V. Dubonos, and A. A. Firsov,
Nature {\bf 438}, 197 (2005).
%
\bibitem{pkim}
	Y. Zhang, Y.-W. Tan, H.L. Stormer, and P. Kim, Nature {\bf 438}, 201 (2005).

\bibitem{graphenereviews}
A. H. Castro Neto, F. Guinea, N. M. R. Peres, K. S. Novoselov and A. K. Geim,
Rev. Mod. Phys. {\bf 81}, 109 (2009);
A. K. Geim and K. S. Novoselov et al., 
Nature Materials {\bf 6}, 183  (2007);
A. K. Geim and A. H. MacDonald,
Physics Today {\bf 60}, 35 (2007).


\bibitem{coey}
Coey et al. Nature 420, {\bf 158} (2002);
El Goresy et al., Science 161, {\bf 363} (1968);
Smith et al, Science 216, {\bf 984} (1982);
Li et al., Applied Phys. Lett. {\bf 90}, 232507 (2007);
P.M. Allemand, Science 253, {\bf 301} (1991);
Esquinazi et al., Phys. Rev. Lett. {\bf 91}, 227201 (2003);
Barzola-Quiquia et al. Phys. Rev. B {\bf 76}, 161403 (2007);
Ohldag et al, Phys. Rev. Lett. {\bf 98}, 187204 (2007);
Cervenka et al., Nature Physics {\bf 5}, 840 (2009);
T. Makarova and F. Palacio, 
`Carbon Based Magnetism: An Overview of the Magnetism of Metal Free Carbon-based Compounds and Materials', Elsevier (2006).


\bibitem{kane}
C. L. Kane and E. J. Mele,
Phys. Rev. Lett. {\bf 95}, 226801 (2005).

\bibitem{arikawa}
M. Arikawa, Y. Hatsugai, and H. Aoki,
Phys. Rev. B {\bf 78}, 205401 (2008).

\bibitem{qiao}
Z. Qiao, S. A. Yang, W. Feng, W.-K. Tse, J. Ding, Y. Yao, J. Wang, and Qian Niu,
Phys. Rev. B 82, 161414(R) (2010).


\bibitem{zhang}
F. Zhang, J. Jung, G. A. Fiete, Q. Niu, A. H. MacDonald, 
arXiv:1010.4003 (2010).

\bibitem{yazyev3}
O. V. Yazyev,
Rep. Prog. Phys. {\bf 73}, 056501 (2010).

\bibitem{alicea}
J. Alicea and M. P. A. Fisher, Phys. Rev. B {\bf 74}, 075422 (2006);
Solid State Comm. {\bf 143}, 504 (2007).

\bibitem{yazyev1} 
O. V. Yazyev, M. I. Katsnelson
Phys. Rev. Lett. {\bf 101}, 037203 (2008).

\bibitem{bhowmick}
S. Bhowmick and V. B. Shenoy,
J. Chem. Phys. {\bf128}, 244717 (2008).

\bibitem{wunsch}
B. Wunsch, T. Stauber, F. Sols, and F. Guinea,
Phys. Rev. Lett. {\bf 101}, 036803 (2008).

\bibitem{monolayer_hf}
J. Jung and A. H. MacDonald, to be published.

\bibitem{joaquin1}  J. Fern\'andez-Rossier and J. J. Palacios,
Phys. Rev. Lett. {\bf 99}, 177204 (2007).

\bibitem{julio}
J. J. Palacios, J. Fern\'andez-Rossier, and L. Brey, Phys. Rev. B {\bf 77}, 195428 (2008).


\bibitem{perdew}
J. P. Perdew and Y. Wang,
Phys. Rev. B {\bf 45}, 13244 (1992).   

\bibitem{pbe}
J. P. Perdew, K. Burke, and M. Ernzerhof,
 Phys. Rev. Lett. {\bf 77}, 3865 (1996). 
  
\bibitem{ceperley}
D. M. Ceperley and B. J. Alder,
Phys. Rev. Lett. 45, 566 - 569 (1980).

\bibitem{wannier}
M. Hatanaka,
Chem. Phys. Lett. {\bf 484}, 276, (2010).

\bibitem{malysheva} 
L. Malysheva and A. Onipko,
Phys. Rev. Lett. 100, 186806 (2008);
arXiv:0802.1385v1 (2008).


\bibitem{ostlund}
A. Szabo and N. Ostlund,
`Modern Quantum Chemistry: Introduction to Advanced Electronic Structure Theory', Dover (1996).


\bibitem{lowdin2} 
P.-O. Lowdin,
Phys. Rev. 97, 1490 (1955).

	
\bibitem{semiinf}
B. Wunsch, T. Stauber, F. Sols, and F. Guinea,
Phys. Rev. Lett. {\bf 101}, 036803 (2008).

\bibitem{zarea}
M. Zarea and N. Sandler,
Phys. Rev. Lett. {\bf 99}  256804 (2007).

\bibitem{gogolin}
R. Egger and A. Gogolin, 
Phys. Rev. Lett. {\bf 79}, 5082 (1997).


\end{thebibliography}
\end{document}